\documentclass[a4paper]{article}
\usepackage[T1]{fontenc}
\usepackage[latin1]{inputenc}
\usepackage{amsmath,amsthm,amsfonts,amssymb, euscript, stmaryrd, graphicx,superbox}
\usepackage{yhmath} 
\usepackage{algorithm}
\usepackage{algorithmic}
\usepackage{color}
\usepackage{listings}
\usepackage{pstricks}
\usepackage{psfrag}
\usepackage{amsthm}
\usepackage{fullpage}
\definecolor{colKeys}{rgb}{0,0,1}
\definecolor{unidentified}{rgb}{0,0,0}
\definecolor{concealments}{rgb}{0,0.5,0}
\definecolor{colString}{rgb}{0.6,0.1,0.1}

\definecolor{Grey1}{rgb}{0.5,0.5,0.5} 
\definecolor{Grey2}{rgb}{0.8,0.8,0.8}
\definecolor{Grey3}{rgb}{0.9,0.9,0.9}
\definecolor{Grey4}{rgb}{0.95,0.95,0.95}
\definecolor{ACORRIGER}{rgb}{0.9,0,0}

\def\myqed {{
\parfillskip=0pt 
\widowpenalty=10000 
\displaywidowpenalty=10000 
\finalhyphendemerits=0 
\leavevmode 
\unskip 
\nobreak 
\hfil 
\penalty50 
\hskip.2em 
\null 
\hfill 
$\square$
\par}} 

\newcommand{\footnoteremember}[2]{
 \footnote{#2}
 \newcounter{#1}
 \setcounter{#1}{\value{footnote}}
}

\author
{
Sylvain \textsc{Corlay}\footnote{Natixis, Equity Derivatives and Arbitrage. E-mail: sylvain.corlay@gmail.com.}
\footnoteremember{myfootnote}{Laboratoire de Probabilités et Modèles Aléatoires, UMR 7599, Université Paris 6, case 188, 4, pl. Jussieu, F-75252 Paris Cedex 5, France.}\\
}

\title{A fast nearest neighbor search algorithm based on vector quantization}
\date{May 24, 2011}

\newtheorem{theo}{Theorem}[section]
\newtheorem{prop}[theo]{Proposition}
\newtheorem{coro}[theo]{Corollary}
\newtheorem*{remark}{Remark}
\newtheorem{lemm}[theo]{Lemma}
\newtheorem{defi}{Definition}

\newcommand{\1}{\textbf{1}}
\newcommand{\N}{\mathbb{N}}

\newcommand{\R}{\mathbb{R}}

\newcommand{\E}{\mathbb{E}}

\newcommand{\PP}{\mathbb{P}}

\newcommand{\supp}{\operatorname{supp}}
\newcommand{\card}{\operatorname{card}}

\newcommand{\var}{\operatorname{Var}}
\newcommand{\Proj}{\operatorname{Proj}}
\newcommand{\cell}{\operatorname{cell}}

\newcommand{\slab}{\operatorname{slab}}

\def\keywordname{{\bf Keywords:}} 
\newcommand{\keywords}[1]{\par\addvspace\baselineskip\noindent\keywordname\enspace\ignorespaces#1}

\graphicspath{{./}{figures/}}
\begin{document}
\lstset{language=c++}
\maketitle
\begin{abstract}
\par In this article, we propose a new fast nearest neighbor search algorithm, based on vector quantization. Like many other branch and bound search algorithms \cite{BentleyKdTree1,McNamesPAT}, a preprocessing recursively partitions the data set into disjointed subsets until the number of points in each part is small enough. In doing so, a search-tree data structure is built. This preliminary recursive data-set partition is based on the vector quantization of the empirical distribution of the initial data-set. 
\par Unlike previously cited methods, this kind of partitions does not a priori allow to eliminate several brother nodes in the search tree with a single test. To overcome this difficulty, we propose an algorithm to reduce the number of tested brother nodes to a minimal list that we call ``friend Voronoi cells''. The complete description of the method requires a deeper insight into the properties of Delaunay triangulations and Voronoi diagrams. 
\end{abstract}
\lstset{language=c++}
\pagebreak
\section*{Introduction}
\par The problem of nearest neighbor search, also known as the post office problem \cite{KnuthArt3} has been widely investigated in the area of computational geometry. It is encountered for many applications, as pattern recognition and vector quantization. 
\par The post-office problem has been solved near optimally for the case of low dimensions. Algorithms differ on their practical efficiency on real data sets. For large dimensions, most solutions have a complexity that grows exponentially with the dimension, or require a bigger query time than the obvious brute force algorithm. In fact, it has been noticed that, if $n$ is the size of the data set and $d$ is the dimensionality, the best choice becomes linear search when $d> K \log(n)$ for some positive constant $K$ which depends on the chosen algorithm. This effect is known as the curse of dimensionality. 
\par As concerns the application to (Voronoi) vector quantization, nearest neighbor projections are recognized to represent the critical part of most codebook optimization algorithms. In this case, the big amount of nearest neighbor searches we have to do shows that a preprocessing of the data-set will be profitable if it reduces the average query time. Still, in some particular cases, the codebook is chosen so that nearest neighbor search is performed easily, (as when dealing with product quantization). Moreover, non-Voronoi quantization methods can also be designed in order to simplify the projection procedure while preserving some important properties of optimal quantizers, as the stationarity in the quadratic case. 
\par Let us also point out that a field recently emerged under the name of dual quantization \cite{DualQuantizationPagesWilbertz1,DualQuantizationPagesWilbertz2}. In this context, the nearest neighbor search, i.e. the location of a point in a Voronoi partition, is replaced by the analogous procedure in the Delaunay triangulation. This localization procedure in Delaunay triangulations have been widely investigated in the practical viewpoint in terms of reduction of its computational complexity. We refer to Devillers, Pion and Teillaud for a review on this subject \cite{TriangulationWalking}.
\par Many nearest neighbor search algorithms rely on a recursive partitioning of the data-set resulting in a search-tree data structure \cite{BentleyKdTree1,McNamesPAT}. The method proposed by McNames in \cite{McNamesPAT} improved the classical Kd-tree algorithm \cite{BentleyKdTree1} by taking advantage of the shape of the data-set thanks to principal component analysis. The ``principal axis tree'' algorithm performs much faster than the classical Kd-tree when the coordinates of the data-set are correlated and it seems to take better the growth of dimensionality. 
\par In our case, the proposed algorithm uses vector quantization as a clustering method to perform this recursive partitioning and to take advantage of the geometry of the data-set. It is classical background that when dealing with empirical distributions, the quadratic vector quantization problem is equivalent to the reduction of the intraclass inertia of the related partition, and the specification of the classical Lloyd algorithm to this case turns out to be the $k$-means clustering algorithm. 
\par We will see that one draw-back of this kind of partition is that, as other tree-based search algorithms, after determining the closest neighbor of a query in a leaf-node of the tree, the procedure has to move up to the parent node and determine whether brother nodes have to be explored or not. Unlike Kd-tree and ``principal axis tree'', our so-called ``quantization tree'' can't eliminate several brother nodes by with a single test. This is the motivation for the development of our friend node algorithm.
\par The paper is organized as follows. Section \ref{sec:voronoi_quantization} is devoted to classical definitions and notations related to vector quantization. The link with the classification problem is pointed out. Section \ref{sec:geometry} recalls in mind some definitions of computational geometry which will be useful in the sequel. As both the fields of vector quantization and algorithmic geometry deal with the notion of Voronoi diagram, we apply ourselves to distinguish the corresponding definitions and notations. Section \ref{sec:kd_and_pat} makes a brief presentation of both the Kd-tree \cite{BentleyKdTree1} and ``principal axis tree'' \cite{McNamesPAT} algorithms. We deal with some optimizations that will be applicable with our quantization tree. Section \ref{sec:quantization_tree} presents the ``crude'' quantization tree, i.e. without using any friend node algorithm. It is presented as the natural counterpart these two branch and bound algorithms with a quantization based partition of the data-set. Section \ref{sec:friend_node_algo} presents the friend node algorithm which was discussed above. Finally, the last section provides some performance comparisons between the different algorithms on various data-sets. 
\section{Vector quantization and Voronoi tessellations}\label{sec:voronoi_quantization}
\par \noindent We consider $(\Omega,\mathcal{A},\PP)$ a probability space and $E$ a (real) finite dimensional Euclidean space. The principle of a random variable $X$ taking its values in $E$ is to approach $X$ by a random variable $Y$ taking a finite number of values in $E$. 
\begin{defi}[quantizer]
	\par \noindent In this surrounding, the discrete random variable $Y$ is a \emph{quantizer} of $X$. 
\end{defi}
\par \noindent If $X \in L^p$, the quantization error is the $L^p$ norm of $|X-Y|$, where $|\cdot|$ denotes the Euclidean norm on $E$. The minimization of this error yields the following minimization problem
\begin{equation}\label{eq:minimization_problem}
\min\{ \|X-Y\|_p, Y:\Omega \to E \textrm{ measurable }, \card(Y(\Omega)) \leq N \}.
\end{equation}
\begin{defi}[Voronoi partition]\label{def:voronoi_quantif}
Consider $N \in \N^*$, $\Gamma = \{\gamma_1,\cdots,\gamma_N\} \subset E$ and let $C = \{C_1,\cdots,C_N\}$ be a Borel partition of $E$. $C$ is a Voronoi partition associated with $\Gamma$ if $\forall i \in \{1,\cdots,N \}, \ C_i \subset \{\xi \in E, |\xi -\gamma_i | = \min\limits_{j \in \{ 1, \cdots, N \} } | \xi - \gamma_j | \}$. 
\end{defi}
\par \noindent If $C = \{C_1,\cdots,C_N\}$ is a Voronoi partition associated with $\Gamma = \{ \gamma_1,\cdots,\gamma_N\}$, it is clear that $\forall i \in \{ 1, \cdots, N \}, \gamma_i \in C_i$. $C_i$ is called Voronoi slab associated with $\gamma_i$ in $C$ and $\gamma_i$ is the center of the slab $C_i$. 
\par \noindent We denote $C_i=\slab_C(\gamma_i)$. For every $a\in \Gamma$, $W(a|\Gamma)$ is the closed subset of $E$ defined by $W(a|\Gamma) = \left\{ y \in E, |y - a | = \min\limits_{\gamma \in \Gamma} |y-\gamma| \right\}.$
\begin{defi}[Nearest neighbor projection]
\par \noindent Consider $\Gamma \subset E$ a finite subset of $E$. A nearest neighbor projection onto $\Gamma$ is an application $\Proj_\Gamma$ that satisfies
$$
\forall x \in E, \quad \big|x - \Proj_\Gamma(x) \big| = \min\limits_{\gamma \in \Gamma} |x-\gamma|.
$$
\end{defi}
\par \noindent To be more precise, if $\Proj_\Gamma$ is a measurable nearest neighbor projection onto $\Gamma$, there exists a Voronoi partition $C = \{C_1,\cdots,C_N \}$ associated to $\Gamma$ such that $\Proj_\Gamma = \sum\limits_{i=1}^N \gamma_i \1_{C_i}$. 
\begin{prop}\label{thm:best_nnp}
	\par \noindent Let $X$ be an $E$-valued $L^p$ random variable, and $Y$ taking its values in the settled point set $\Gamma = \{y_1,\cdots,y_N\} \subset E$ where $N \in \N$. Set $\widehat{X}^\Gamma$ the random variable defined by $\widehat{X}^\Gamma := \Proj_{\Gamma}(X)$ where $\Proj_{\Gamma}$ is a nearest neighbor projection on $\Gamma$, called a Voronoi $\Gamma$-quantizer of $X$. 
	\par \noindent Then we clearly have $\left|X-\widehat{X}^\Gamma\right| \leq |X-Y| \textrm{ a.s.}$. Hence $\left\| X-\widehat{X}^\Gamma \right\|_p \leq \| X-Y\|_p$.
\end{prop}
\par \noindent A consequence of this proposition is that solving the minimization problem (\ref{eq:minimization_problem}) amounts to solving the simpler minimization problem
\begin{equation}\label{eq:simpler_minimization_problem}
\min \left\{ \| X-\Proj_{\Gamma}(X) \|_p, \ \Gamma \subset E, \card(\Gamma) \leq N\right\}.
\end{equation}
\par \noindent The quantity $\left\| X-\Proj_{\Gamma}(X) \right\|_p$ is called the mean $L^p$-quantization error. When this minimum is reached, we refer to $L^p$-optimal quantization. 
\par \noindent The problem of the existence of a minimum have been investigated for decades on its numerical and theoretical aspects in the finite dimensional case \cite{PagesIntegVectorQuant,GrafLushgyMonograf}. For every $N \geq 1$, the $L^p$-quantization error is Lipschitz-continuous and reaches a minimum. An $N$-tuple that achieves the minimum has pairwise distinct components, as soon as $\card(\supp(\PP_X)) \geq N$. This result stands in the general case of a random variable valued in a reflexive Banach space \cite{LuschgyPagesFunctional3}. If $\card(X(\Omega))$ is infinite, this minimum strictly decreases to $0$ as $N$ goes to infinity. The asymptotic rate of convergence, in the case of non singular distributions is ruled by the Zador theorem \cite{GrafLushgyMonograf}. A non-asymptotic upper bound for the quantization error is also available \cite{FunctionalQuantizationLevy}. 
\par \noindent We now focus on the quadratic case ($p=2$). For a $L^2$ random variable $X$, we now denote $\mathcal{C}_N(X)$ the set of $L^2$-optimal quantizers of $X$ of level $N$ and $e_N(X)$ the minimal quadratic distortion that can be achieved when approximating $X$ by a quantizer of level $N$. A quantizer $Y$ of $X$ is stationary (or self-consistent) if 
$Y = \E[X|Y]$. 
\begin{prop}[Stationarity of $L^2$-optimal quantizers]
A (quadratic) optimal quantizer is stationary. 
\end{prop}
\par \noindent The stationarity is a particularity of the quadratic case. In other $L^p$ cases, a similar property involving the notion of $p$-center occurs. A proof is available in \cite{GLPApprox}.
\begin{defi}[Centroidal projection]\label{def:centroidal_projection}
	\par Let $C = \{C_1,\cdots,C_N\}$ be a Borel partition of $E$. Let us define for $1\leq i\leq N$, 
 	$G_i = 	\left\{\begin{array}{llll}
 			\E[X | X \in C_i] & \textrm{if } \PP[X \in C_i] \neq 0,\\
 			0 & \textrm{in the other case,}
 			\end{array}\right.$
 	the centroids associated with $X$ and $C$. 
 	\par The centroidal projection associated $C$ and $X$ is the application $\Proj_{C,X} : x \to \sum\limits_{i=1}^N G_i \1_{C_i}(x)$.
\end{defi}

\begin{lemm}[Huyghens, variance decomposition]\label{thm:huyghens_quantif}
\par Let $X$ be a $E$-valued $L^2$ random variable, $N \in \N^*$ and $C=(C_i)_{1 \leq i \leq N}$ a Borel partition of $E$. Consider $\Proj_{C,X} = \sum\limits_{i=1}^N G_i \1_{C_i}$ the associated centroidal projection. Then one has, 
$$
\var(X) = \underbrace{\E\left[\left|X-\Proj_{C,X}(X)\right|^2\right]}_{:=(1)}+\underbrace{\E\left[\left|\Proj_{C,X}(X)-\E[X]\right|^2\right]}_{:=(2)}.
$$
\par \noindent The variance of the probability distribution $X$ decomposes itself as the sum of the \textbf{intraclass inertia} $(1)$ and the \textbf{interclass inertia} $(2)$.
\end{lemm}
\par \noindent \textbf{Proof:}
\vspace{-\baselineskip }
$$
\begin{array}{lll}
\quad \var(X) & = \E\left[\left|X-\Proj_{C,X}(X) + \Proj_{C,X}(X) - \E[X]\right|^2\right] \\
		& = \underbrace{\E\left[\left|X- \Proj_{C,X}(X)\right|^2\right]}_{=(1)} + \underbrace{\E\left[\left|\Proj_{C,X}(X)-\E[X]\right|^2\right]}_{=(2)}\\
		& \qquad \qquad \qquad \qquad \qquad \qquad + \underbrace{2\E\left[\left\langle X-\Proj_{C,X}(X),\Proj_{C,X}(X)-\E[X]\right\rangle\right]}_{:=(3)}.
\end{array}
$$
Now $(3) = 0$ since $\Proj_{C,X}(X) = \E\left[X\middle|\Proj_{C,X}(X)\right]$. \myqed
\section{Backgrounds on theory of polytopes}\label{sec:geometry}
\par Let $E$ be a $d$ dimensional vector space and $E^*$ its dual. 
\begin{defi}[$k$-flat]
	\par \noindent A $k$-flat is a $k$-dimensional affine subspace $E$.
\end{defi}
\begin{defi}[convex polyhedron and convex polytope]
	\par \noindent A convex polyhedron is the intersection of a finite subset of closed halfspaces. If it is bounded, it is a convex polytope. 
\end{defi}
\begin{defi}[cell]
\par \noindent A cell is the intersection of a finite set of flats and open halfspaces. And thus, equivalently, it is the relative interior of a convex polyhedron. If $R \subset E$, we denote $\cell(R)$ the relative interior of the convex hull of $R$.
\end{defi}
\begin{defi}[simplex]
	\par \noindent A simplex is $\cell(R)$ where $R$ is a set of affinely independent points.
\end{defi}
	\begin{itemize}
		\item A $2$-dimensional simplex is the interior of a triangle.
		\item A $3$-dimensional simplex is the interior of a tetrahedron.
	\end{itemize}
\begin{defi}[circumsphere]
	\par \noindent A circumsphere of a set $R \subset E$ is a sphere $S$ of $E$ such that $R \subset S$.
\end{defi}
\begin{defi}[supporting halfspace]
	\par \noindent Let $C$ be a convex subset of $E$. A hyperplane $H$ supports $C$ if $H\cap C\neq \emptyset$ and $C$ is contained into one of the closed halfspaces defined by $H$.
\end{defi}
\begin{lemm}
	\par \noindent Let $C \varsubsetneq E$ be a convex subset of $E$. If $H$ is a supporting hyperplane of $C$, then every point of $H\cap C$ is a frontier point of $C$.
\end{lemm}
\par \noindent \textbf{Proof:} Let $H$ be a supporting hyperplane of $C$ of equation $\phi(x)=\alpha$. Consider $v \in E$ such that $\forall x \in E \ \phi(x) = \langle x |v \rangle$. 
\par \noindent Consider $a \in H\cap C$. We may assume that $\forall x \in C \ \phi(x) = \langle x |v \rangle \geq \alpha$. If $a$ does not belong to the boundary of $C$, $\exists \varepsilon \geq 0, B(a, \varepsilon)\subset C$ so for any $\lambda >0$ small enough, $a-\lambda v \in C$ and
$$
\alpha \leq \phi(a-\lambda v) = \langle a | v \rangle - \lambda \|v\|^2 < \langle a | v \rangle = \alpha
$$
which yields a contradiction. Consequently $a\in\partial C$. \myqed
\begin{coro}
\par \noindent Every point of the boundary of a convex subset of $E$ belongs to one of its supporting hyperplanes.
\end{coro}
\par \noindent \textbf{Proof:} The proof is straightforward using the same approach as for the previous lemma. \myqed
\begin{lemm}
\par If $C$ is a non empty closed convex subset of $E$, distinct of $E$, then every point of the boundary $\partial C$ belongs to a supporting hyperplane of $C$.
\end{lemm}
\par \noindent \textbf{Proof:} $a \in \partial C \Rightarrow \forall k \in \N^*, \exists x_k \in B\left(a,\frac{1}{k}\right), x_k \notin C$. We denote $y_k=p_C(x_k)$ the projection of $x_k$ on $C$, $z_k = \frac{x_k-y_k}{\|x_k-y_k\|}$. Owing to the characterization of the projection on a closed convex subset, we have 
\vspace{-0.6\baselineskip}
$$
\begin{array}{lll}
\forall z \in C, \ \langle x_k-p_C(x_k), x_k-z \rangle 	&= |x_k-p_C(x_k)|^2- \overbrace{\langle x_k-p_C(x_k), z-p_C(x_k) \rangle}^{\leq 0}\\
															&\geq |x_k-p_C(x_k)|^2 >0 \hspace{3mm} \textrm{because} \hspace{3mm} x_k\notin C.
\end{array}
$$
\par \noindent Every vector $z_k$ lying on the unit sphere of $E$ (which is compact), one can extract a subsequence of $z_{\phi(k)}$ that converges to a vector $v$, with $|v|=1$. As $(x_k)_{1\leq k }$ converges to $a$, by continuity of $p_C$ and of the scalar product, we have
$$
\forall z \in C, \langle v, a-z \rangle = \lim\limits_{k \to +\infty} \langle z_{\phi(k)},x_{\phi(k)}-z \rangle \geq 0.
$$
\par \noindent In other words $C$ is contained in the halfspace $\{ z \in E, \langle v,a-z \rangle \geq 0\}$. Moreover, as $a$ is in the corresponding hyperplane $H$, $H$ is a supporting halfspace of $C$. \myqed
\begin{defi}[face]
	\par A face of a convex polyhedron $P$ is the relative interior of the intersection of a hyperplane supporting $P$ with the closure of $P$.
\end{defi}
\begin{prop}
	\par Let $P$ be a convex polyhedron, a face of $P$ is a cell, and a face of a face of $P$ is a face of $P$.
\end{prop}
\begin{defi}[$k$-face]
	\par A $k$-face is a a face whose affine closure has dimension $k$.
\end{defi}

\begin{defi}[cell complex]
	\par \noindent A cell complex is a finite collection of pairwise disjoint cells so that the face of every cell is in the collection. 
\end{defi}
\begin{defi}[opposite $k$-faces]
	\par \noindent Two distinct $k$-cells of a cell complex are opposite if they have a common $(k-1)$-face. 
\end{defi}
\begin{defi}[triangulation]\label{def:triangulation}
	\par \noindent Let $S$ be a finite point set of $E$. A triangulation $T$ of $S$ is a cell complex whose union is the convex hull of $S$ and whose set of $0$-cells is $S$.
\end{defi}
\par \noindent Definition \ref{def:triangulation} is a non standard definition because cells are not required to be simplices. This formalism is due to Steven Fortune \cite{FortuneVoronoiDelaunay}.
\begin{defi}[proper triangulation]
\par \noindent A proper triangulation is a triangulation whose all cells are simplices. 
\end{defi}
\par \noindent Any triangulation can be completed to a proper triangulation by subdividing non simplicial cells. 
\subsection{Voronoi diagrams and Delaunay triangulations}\label{sec:computational_geometry}
\subsubsection*{Voronoi diagram}
\par \noindent Let $E$ be a $d$-dimensional Euclidean space, and $S$ a finite subset of $E$. In the following, elements of $S$ will be called \emph{sites}.
\begin{defi}[Voronoi cell]\label{defi:voronoi_cell}
	\par \noindent For a nonempty subset of $S$, $R \subset S$, the Voronoi cell of $R$, denoted $V(R)$ is the set of all points in $E$ that are equidistant from all sites in $R$, and closer to every site of $R$ than to any site not in $R$.
\end{defi}
\begin{prop}\label{thm:slab_and_cells}
\begin{itemize}
\item Clearly, is $r \in S$, $V(\{r\})$ is the set of all points strictly closer to $r$ than to any other site. In particular, it is the interior of the Voronoi slab associated to $r$ in $S$. (See the definition of a Voronoi slab in Section \ref{sec:voronoi_quantization}.) 
\item $V(R)$ may be empty.
\item Any point of $E$ lies in $V(R)$ for some $R\subset S$.
\end{itemize}
\end{prop}
\begin{defi}[Voronoi diagram]
\par \noindent The Voronoi diagram $V$ is the collection of all nonempty Voronoi cells $V(R)$ for $R\subset S$. 
\end{defi}
\subsubsection*{Delaunay triangulation}
\begin{defi}[Delaunay cell]
\par \noindent If $R\subset S$, and $V(R)$ is a non empty Voronoi cell, then the Delaunay cell $D(R)$ is $\cell(R)$. 
\end{defi}
\begin{defi}[Delaunay triangulation]
\par \noindent The Delaunay triangulation $D$ of $S$ is the collection of Delaunay cells $D(R)$, where $R$ varies over subsets of $S$ with $V(R)$ non empty.
\end{defi}
\begin{prop}[Empty circumsphere property]\label{thm:empty_circumsphere}
\par For $R\subset S$, $\cell(R)$ is a Delaunay cell if and only if there is is a circumsphere of $R$ that contains no site of $S \backslash R$ in its interior.
\end{prop}
\par \noindent \textbf{Proof:} Such a circumsphere can be obtained with center an point in the Voronoi cell $V(R)$. \myqed
\begin{figure}[!ht]
	\hspace{.17\linewidth}
	\psfrag{Voronoidiagram}{\small{Voronoi diagram}}	
	\psfrag{Delaunaytriangulation}{\small{Delaunay triangulation}}	
	\psfrag{Dataset}{\small{Data set $S$}}	
	\psfrag{$s_1$}{\footnotesize{$s_1$}}	
	\psfrag{$s_2$}{\footnotesize{$s_2$}}	
	\psfrag{$s_3$}{\footnotesize{$s_3$}}	
	\psfrag{$c_123$}{\footnotesize{$C_{123}$}}
	\psfrag{$C$}{\footnotesize{$C$}}	
	\psfrag{$-2$}{\footnotesize{$-2$}}	
	\psfrag{$-1.5$}{\footnotesize{$-1.5$}}
	\psfrag{$-1$}{\footnotesize{$-1$}}
	\psfrag{$-0.5$}{\footnotesize{$-0.5$}}		
	\psfrag{$0$}{\footnotesize{$0$}}
	\psfrag{$0.5$}{\footnotesize{$0.5$}}		
	\psfrag{$1$}{\footnotesize{$1$}}
	\psfrag{$1.5$}{\footnotesize{$1.5$}}	
	\psfrag{$2$}{\footnotesize{$2$}}
	\includegraphics[height=7cm]{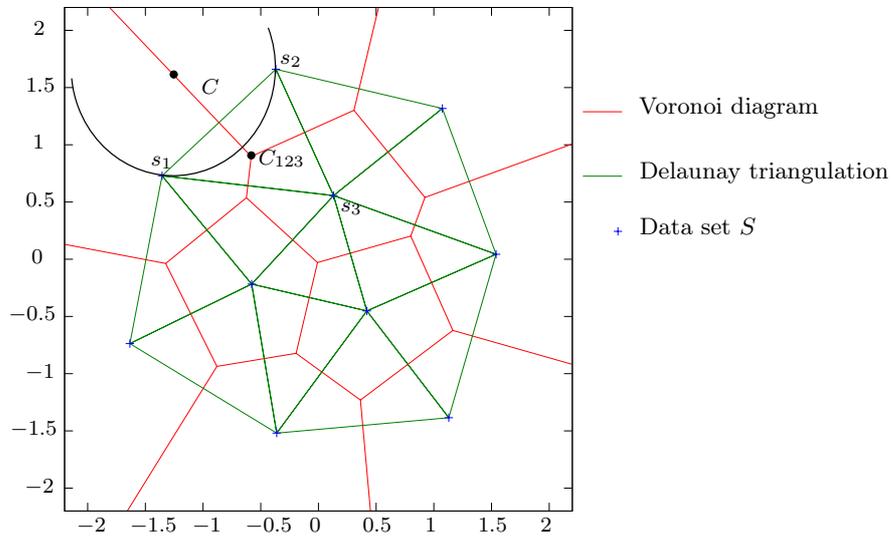}
	\caption{Voronoi diagram and Delaunay triangulation of a data set $S$ of size $10$. We have $C \in V_S(\{s_1,s_2\})$. So $C$ is the center of an empty circumsphere of $\{s_1,s_2\}$. The point $C_{123}$ is the center of the circumsphere of the Delaunay triangle $\{ s_1,s_2,s_3 \}$.} 
\end{figure}
\begin{theo}\label{thm:unbounded_voronoi_cell}
	\par Let $S$ be a set of $n$ points in $E$ with Voronoi diagram $V$ and Delaunay triangulation $D$. Then 
	\begin{enumerate}
	\item $V$ is a cell complex that partitions $E$.
	\item $D$ is a triangulation of $S$.
	\item $V$ and $D$ are linked with the following duality relation:
		\par For $R,R'\subset S$, $V(R)$ is a face of $V(R')$ if and only if $D(R')$ is a face of $D(R)$.
	\item $V(R)$ is unbounded if and only if every site of $R$ is on the boundary of the convex hull of $S$.
	\end{enumerate}
\end{theo}
\par \noindent We refer to \cite{FortuneVoronoiDelaunay} for a detailed proof.
\subsubsection*{Locality}
\begin{defi}[locally Delaunay]
\par \noindent We consider two opposite $d$-cells $\cell(R)$ and $\cell(R')$ in a triangulation $T$ with circumspheres $C$ and $C'$. $\cell(R)$ and $\cell(R')$ are locally Delaunay if $R'\backslash R$ is outside of $C$. This is equivalent to $R\backslash R'$ outside of $C'$. 
\par \noindent A triangulation is locally Delaunay if every pair of opposite $d$-cells is locally Delaunay.
\end{defi}
\begin{lemm}[Delaunay and locally Delaunay]\label{thm:locality_delaunay}
\par \noindent A triangulation is Delaunay if and only if it is locally Delaunay.
\end{lemm}
\par \noindent We refer to \cite{FortuneVoronoiDelaunay} for a detailed proof.

\begin{defi}[General position]
Let $S$ be a nonempty finite set of sites in $E$. $S$ is in general position if no $d+1$ points of $S$ are affinely dependent and if no $d+2$ points of $S$ lie on a common sphere. 
\end{defi}
\begin{defi}[Incircle list]
\par In the following, if $S$ is a finite nonempty set of sites, $D$ is a Delaunay triangulation of $S$ and $x \in E$ is a settle point, we call incircle list and denote $ICL_D(x)$ the set of $d$-cells of $D$ whose circumsphere contains $x$. 
\end{defi}
\par \noindent If $S$ is in general position, no Delaunay cell of $S$ is degenerate. Every cell of the triangulation is a simplex and for any $R \subset S$, $V(R)$ has dimension $d+1 - |R|$.
\subsubsection*{Computing the Delaunay triangulation and the Voronoi diagram}\label{sec:voronoi_delaunay_algo}
\par Whereas the Voronoi diagram was defined before the Delaunay triangulation, it has been recognized that it is easier to devise algorithms in terms of Delaunay triangulation, especially because of the locality property \ref{thm:locality_delaunay}. 
\par A common data structure for Delaunay triangulations is a graph structure where each simplex is a ``node''. The node contains the indices of the $d+1$ sites of the simplex and the pointers to the adjacent simplices. Null pointers are used when the simplices lie on the boundary of the triangulation. Cells of lower dimension are not directly represented in the graph structure. Another convenient convention is that the $k$th pointer stored in the node corresponds to the facet obtained by deleting the  $k$th site in the node. Moreover the order is chosen so that the orientation of every simplex in the triangulation remains always positive. 
\par Here, we present the principles of incremental algorithms for Delaunay triangulations. In this kind of algorithms, sites are added one by one, and the Delaunay triangulation is modified to include each new site. Many other algorithms have been designed for computing the Delaunay triangulation, especially in dimension $2$. Moreover, computing the Delaunay triangulation of the Voronoi diagram in the one-dimensional case simply amounts to sorting the data set. An advantage of incremental algorithms is that they are valid in any dimension. Moreover, for another purpose in the following, we will need a new algorithm (the friend node algorithm presented in Section \ref{sec:friend_node_algo}) that requires a stage which is very similar to the insertion of a new point in the Delaunay triangulation. Hence we will focus here on incremental algorithms. 
\par \noindent Let $S = (s_1,\cdots,s_N)$ be a nonempty finite set of sites of $E$ of cardinal $N$. We define the sets $S_k:=(s_1,\cdots,s_k)$ for $k \in \{1,\cdots,N\}$. Now, for a settled $i<N$, let us consider $D_i$ the Delaunay triangulation of $S_i$. We inspect the situation of $s_{i+1}$ with respect to the Delaunay triangulation $D_i$. From this analysis, the Delaunay triangulation will be modified locally to build a new Delaunay triangulation $D_{i+1}$ of $S_{i+1}$. When all the sites of $S$ will be processed, we will have the complete Delaunay triangulation $D$ of $S$. 
\vspace{2mm}
\par \noindent Three situations can occur, if $S$ is in general position:
\begin{enumerate}
\item $s_{i+1}$ lies in the interior convex hull of $S_i$. 
\item $s_{i+1}$ does not lie in any circumsphere of any simplex of $D_i$.
\item $s_{i+1}$ lies outside of the convex hull of $S_i$ but belongs to a circumsphere of a simplex of $D_i$.
\end{enumerate}
\par \noindent $(1)$ In the first situation, let denote $\mathcal{S}:= ICL_{D_i}(s_{i+1})$ and $F_1,\cdots, F_p$ the external faces of $\mathcal{S}$ of any dimension $k<d$. We can show that the cell complex defined by
$$
D_{i+1} := (D_i\backslash \mathcal{S}) \ \cup \big\{ \cell(F_j,s_{i+1})_j, 1 \leq j \leq p \big\} \cup \big\{ \left\{ s_{i+1}\right\} \big\}
$$
is the Delaunay triangulation associated to $S_{i+1}$. In a more general setting, we have the following property: 
\begin{prop}[star-shaped incircle list]\label{prop:star_shaped}
\par Let $S$ be a nonempty finite set of sites of $E$ and $x\in E$ that lies on the convex hull of $S$. Consider $C$ the union of the $d$-cells of $ICL_D(x)$ and of all its faces. Then $C$ is star-shaped from $x$, that is for any point $p \in C$, $[x,p] \subset C$.
\end{prop}
\vspace{2mm}
\par \noindent $(2)$ The second situation is the simplest. If $F_1, \cdots, F_p$ are the external faces of the triangulation $D_i$ (of any dimension $k<d$) that are visible from $s_{i+1}$. We can show that the cell complex defined by 
$$
D_{i+1} := D_i \ \cup \big\{ \cell(F_j,s_{i+1})_j, 1 \leq j \leq p \big\} \cup \big\{ \left\{ s_{i+1}\right\} \big\}
$$
is the Delaunay triangulation associated to $S_{i+1}$.
\vspace{2mm}
\par \noindent $(3)$ In the third situation, if we denote $\mathcal{S}=ICL_{D_i}(s_{i+1})$ the set of elements of $D_i$ whose circumsphere contains $s_{i+1}$ and $F_1,\cdots, F_p$ are the external faces) of this set which are not visible from $s_{i+1}$ and $F_{p+1},\cdots,F_{p+q}$ are the external faces of $D_i$ that are not faces of elements of $\mathcal{S}$ and that are visible from $x_{i+1}$. We can show that the cell complex defined by
$$
D_{i+1} := (D_i\backslash \mathcal{S}) \ \cup \big\{ \cell(F_j,s_{i+1})_j, 1 \leq j \leq p \big\} \cup \big\{ \left\{ s_{i+1}\right\} \big\}
$$
is the Delaunay triangulation associated to $S_{i+1}$.
\vspace{2mm}
\par \noindent The first triangulation $D_{d+1}$ is made of a simple simplex defined by the $d+1$ first inserted points. 
\vspace{2mm}
\par One important modification of the incremental algorithm consists in inserting sites in a random order. Its expected running time is better than the worst case running time for the incremental algorithm. 
\par The worst case complexity of computing the Delaunay triangulation of $n$ points in a $d$ dimensional Euclidean space $E$ is $O\left(n \log (n) + n^{\left\lceil \frac{d}{2} \right\rceil}\right)$. 
\vspace{2mm}
\par \textbf{On the practical implementation}
\par \noindent The first step is the Localization. It consists in finding whether the new site $x$ is in the convex hull of $S$ or not, and if it is the case, in what Delaunay cell of the triangulation $T_S$ $x$ lies. A survey on localization methods is available in \cite{TriangulationWalking}. When $x$ is inside of the convex hull of $S$, the localization procedure return the index of the the Delaunay cell where it lies. This corresponds to the situation $(1)$. When $x$ is outside of this convex hull, the localization returns a Null pointer. This corresponds to the situations $(2)$ or $(3)$. 
\vspace{2mm}
\par \noindent The second step consists in finding the list of the Delaunay cells whose circumsphere contains $x$ (the incircle list). In the situation $(1)$, this list contains at least the Delaunay cell where $x$ is located. Owing to the Proposition \ref{prop:star_shaped}, we know that the union of these Delaunay cells is star-shaped so that it can be determined locally by testing connected cells in the graph structure presented above. 
\vspace{2mm}
\par \noindent The last step consists in deleting the Delaunay cells of the incircle list and connecting the new site to the external faces of the incircle list or the visible faces of the convex hull of $S$ depending on the situation $(1)$, $(2)$ or $(3)$.
\section{Classical examples of fast nearest neighbor search algorithms in low dimensions}\label{sec:kd_and_pat}
\par Given a set of $n$ points, $\{x_1,\cdots,x_n\} \subset E$, the nearest neighbor problem is to find the point that is closest to a query point $q \in E$. Many algorithms have been proposed to avoid the large computational cost of the obvious brute force algorithm. When one has to perform a big amount of nearest neighbor searches, a preprocessing of the data set will be profitable if it reduces the average query time. 
\par The problem is optimally solved in the case of dimension $1$, where the best algorithm is, as a preprocessing to sort the data set by the unique coordinate of its points. (Approximative cost of $O(n \ln(n))$). The search algorithm consists of a simple binary search whose cost is $\frac{\ln(n)}{\ln(2)}+O(1)$. 
\par In the case of low dimensions, most fast search algorithms still have an approximative preprocessing cost of $O(n \log(n))$ and an average search cost in $O(\log(n))$ in low dimension. The criterion of choice among them relies on
\begin{itemize}
\item their effective speed on real data sets,
\item the required memory,
\item the sensitivity of the speed to the dimensionality.
\end{itemize}
\vspace{2mm}
\par A first obvious optimization called \emph{partial distance search} (P.D.S.) consists of a simple modification of the brute force search: during the calculation of the distance, if the partial sum of square differences exceeds the distance to the nearest neighbor found so far, the calculation is aborted. This almost always speeds up the nearest neighbor search procedure. 
\subsection{The Kd-tree algorithm}
\par The Kd-tree algorithm is the archetype of the branch-and-bound nearest neighbor search tree. It is very popular because of its simplicity. 
\vspace{2mm}
\par \noindent \textbf{Building the tree:} 
\begin{itemize}
\item Every point of the data set is associated to the root node.
\item The data set is being sorted by its first coordinate. Then it is divided in two subsets of cardinal $\left\lfloor \frac{n}{2} \right\rfloor+1$ or $\left\lfloor \frac{n}{2} \right\rfloor$.
\item Each subset is associated to a child node of the root node.
\item The process is repeated on each child node recursively using the coordinate axis in a cyclic order, until there are less than two points in each node.
\end{itemize}
\par \noindent \textbf{Searching in the tree:} Let $q$ be the query point.
\begin{itemize}
\item The search procedure begins by searching in what child node $q$ is (depending of its first coordinate).
\item This child node is then searched, and the process is repeated recursively until a terminal node is reached.
\item A trivial nearest neighbor search is performed in the terminal node. (Partial Distance Search optimization can be used.)
\item The procedure moves up to the parent of the terminal node.
\item If the distance $d_2$ between $q$ and the hyperplane that splits the data set is smaller than the distance $d_{\min}$ to the nearest neighbor found so far, the other child node is searched.
\item The procedure continues its way back to the root node.
\end{itemize}
\begin{figure}[!ht]
	\begin{center}
	\psfrag{flatex_d2}{\small{$d_2$}}
	\psfrag{flatex_dist}{\small{$d_{\min}$}}
	\psfrag{$q$}{\small{$q$}}
	\psfrag{$p$}{\small{$p$}}
	\includegraphics[width=6cm]{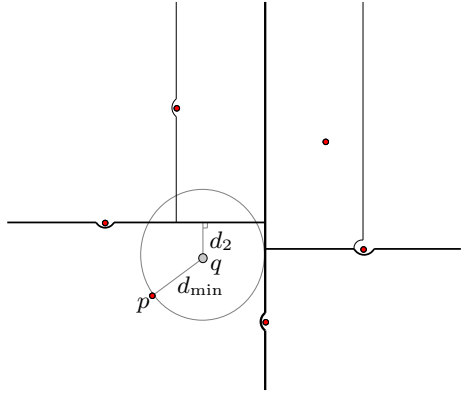}
	\caption{K-d tree elimination condition: if the distance $d_2$ between the query point $q$ and the brother node is smaller than the distance $d_{\min}$ to the nearest neighbor found so far, say $p$, the brother node has to be explored.}
	\label{fig:kd_tree_elimination}
	\end{center}
\end{figure}
\par \noindent \textbf{Complexity:} Except in one dimension where the search complexity is logarithmic (it amounts to a binary search), the worst case of the Kd-tree corresponds to the case where every node of the tree is explored. Then the worst case complexity is time exponential. The distances to every point is computed. The complexity of the preprocessing is $O(d\times n\log(n))$. 

\subsection{The principal axis tree algorithm}\label{sec:PAT}
The Principal Axis Tree (PAT) is a generalization of the Kd-tree proposed by McNames in \cite{McNamesPAT}. Instead of using a coordinate axis to sort the data set, its principal axis is used at each step. Moreover, the number of child node in the tree can be greater than $2$ at each generation. 
\vspace{2mm}
\par \noindent \textbf{Building the tree:}
\begin{itemize}
\item Every point of the data set is associated to the root node.
\item The data set is being sorted by its projection on its principal axis. Then it is partitioned in $n_c$ subsets whose cardinality is $\left\lfloor \frac{n}{n_c} \right\rfloor +1$ or $\left\lfloor \frac{n}{n_c} \right\rfloor$. 
\item Each subset is associated to a child node of the root node.
\item The process is repeated on each child node recursively until there are less than $n_c$ points in each node. 
\item At each step, the principal axis, and maximal and minimal values of subset's projection on the principal axis are kept in memory.
\end{itemize}

\par \noindent \textbf{Optimizing the elimination condition:}

\begin{figure}[!ht]
	\begin{center}
	\psfrag{flatex_region1}{\footnotesize{Region $1$}}
	\psfrag{flatex_region2}{\footnotesize{Region $2$}}
	\psfrag{flatex_region3}{\footnotesize{Region $3$}}
	\psfrag{flatex_region4}{\footnotesize{Region $4$}}	
	\psfrag{flatex_region5}{\footnotesize{Region $5$}}
	\psfrag{flatex_d_{q2}}{\tiny{$d_{q2}$}}	
	\psfrag{flatex_d_{qx}}{\tiny{$d_{qx}$}}	
	\psfrag{flatex_d_{2x}}{\tiny{$d_{2x}$}}	
	\psfrag{flatex_d_{23}}{\tiny{$d_{23}$}}	
	\psfrag{flatex_d_{34}}{\tiny{$d_{34}$}}	
	\psfrag{flatex_q}{\footnotesize{$q$}}	
	\psfrag{flatex_x}{\footnotesize{$x$}}	
	\psfrag{flatex_b_4}{\footnotesize{$b_4$}}
	\psfrag{flatex_b_2}{\footnotesize{$b_2$}}	
	\psfrag{flatex_b_3}{\footnotesize{$b_3$}}	
	\psfrag{flatex_b_4}{\footnotesize{$b_4$}}
	\includegraphics[width=9cm]{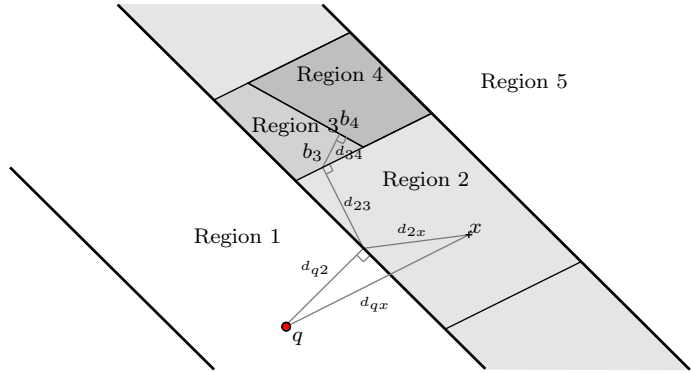}
	\caption{Elimination condition of the principal axis tree.}
	\label{fig:pat_elimination_optimization}
	\end{center}
\end{figure}
\par \noindent We refer here to Figure \ref{fig:pat_elimination_optimization}. We can improve the lower bound to the points that belong to child nodes of brother nodes. For any point $q$ in region $1$ and $x$ in region $2$, we have $d^2(q,x) \geq d^2_{q2} + d^2_{2x}$. This result is then used again to get a lower bound to points in region $3$, and $4$ and so on.
$$
\begin{array}{cccl}
d^2_{2x} & \geq & d^2_{23}				& \forall x \in \textrm{ Region } 3,\\
d^2(q,x) & \geq & d^2_{q2} + d^2_{23} + d^2_{34} & \forall x \in \textrm{ Region } 4.
\end{array}
$$
\par \noindent \textbf{Searching in the tree:} Let $q$ be the query point. 
\begin{itemize}
\item The search process begins by searching in which child node $q$ is (by computing its projection on principal axis).
\item This child node is then searched, and the process is repeated recursively until a terminal node is reached.
\item A partial distance search is then performed in the terminal node.
\item The procedure moves up to the parent of the terminal node.
\item The elimination condition is checked to decide if brother nodes have to be searched or not.
\item The procedure continues its way back to the root node. 
\end{itemize}
\par \noindent \textbf{Choice of parameter $n_c$:} For normal or uniform random data sets (and distribution of query points), best overall performances are obtained with $n_c=7$ (independently from dimensionality for $d<10$). (The same optimal value is obtained by McNames in \cite{McNamesPAT}.)
In the case where the data set is an optimal quantizer of those distributions, best performance is obtained with $n_c = 13$. 
\vspace{2mm}
\par \noindent \textbf{Complexity:} Space storage is $O(n)$. Except in the one-dimensional setting where the search complexity is logarithmic (it comes to a binary search), the worst case of the Kd-tree corresponds to the case where every node of the tree is explored. Then the worst case complexity is time exponential ($2^n$ comparisons of coordinates). $n$ distances are computed. The complexity of the preprocessing is $O(d \times n \log(n))$.
\vspace{2mm}
\par \noindent \textbf{Algorithm performance:} On a $5000$ points Gaussian data set in $\R^2$, the depth of the tree is $4$. 
\begin{itemize}
\item $27$ (partial) distances,
\item $15$ scalar products,
\item $9$ binary searches
\end{itemize}
are performed in average. 
\par \noindent \textbf{Why using this space partitioning ?} The idea is that good empirical performance of PAT are due to the fact that it takes advantage of the shape of the data set. Yet obviously when both query point distribution and data sets lie on a smaller dimension ($k<d$) subspace of $E$, one retrieves the same complexity as when using the same algorithm on a $k$ dimensional space. This intrinsic dimension is often less than the spatial dimension of the space. In a more general setting, PAT takes advantage of high correlations in the data set coordinates. 
\par However if one uses the same number of child nodes $n_c$ in Kd-tree and PAT tree, we see that
\begin{itemize}
\item Preprocessing time is longer for PAT than for Kd-tree. 
\item The first traversal of the tree to a terminal node is more costly (projections have to be computed).
\end{itemize}
\par \noindent But PAT is still faster because its geometrical partition of the space fits the data set in a more relevant way. To be precise, it happens less often than one has to search a brother node with PAT than with Kd-tree. 
\par In \cite{DhaesVanDyckRodetNNS}, the same space decomposition was proposed for the nearest neighbor search problem (but using the only $2$ child node at each generation). They justify the use of this decomposition using a heuristic criterion, according to which the best possible decomposition of the data-set into two subsets for branch and bound nearest neighbor search is to split the data set with respect to its projection on the principal axis. 
\section{A new quantization based tree algorithm}\label{sec:quantization_tree}
\par As we have seen in previous sections, a good space decomposition that fits to the data distribution may lead to a faster branch and bound nearest neighbor search algorithm, if less brother nodes have to be explored. The traversal of the tree can be a little more expensive if it is compensated by the gain due to the fact that less nodes are explored.
\par Principal component analysis and optimal quantization are two types of projection of a probability distribution. Similar inertia decompositions hold in the quadratic case (Huyghens lemma).
\par PAT is based on a recursive space decomposition based on the principal component analysis of the underlying data set. The initial idea here is to design a branch and bound algorithm based on a recursive quantization of the empirical distribution of the underlying data set. 
\subsection{The crude quantization tree algorithm}
\par \noindent \textbf{Building the tree:}
\begin{itemize}
\item Every point of the data set is associated to the root node.
\item The data set is being partitioned into $n_c$ subsets corresponding to the Voronoi cells of an optimized quantizer of the empirical distribution of the data set.
\item Each subset is associated to a child node of the root node.
\item The process is repeated on each child node recursively until there are less than a certain number of points in each node.
\end{itemize}
\par \noindent Some other computations are done during the preprocessing that will be detailed further on.
\begin{remark}
\par One notices that the resulting search tree is not balanced and may have some longer branches.
\end{remark}
\par \noindent \textbf{Searching in the tree:} Let $q$ be the query point.
\begin{itemize}
\item By performing trivial nearest neighbor researches in the node's quantizer the search algorithm traverses the tree to a terminal node where a trivial partial distance search is performed.
\item The procedure moves up to the parent of the terminal node.
\item The elimination condition, (developed further on) is checked to decide whether brother nodes have to be searched or not. 
\item The procedure continues its way back to the root node.
\end{itemize}
\par \noindent \textbf{Consistency of the space decomposition:} 
\par Implementing only the way down to the terminal node (with $n_c = 7$ in both principal axis tree and quantization tree), we naturally do not obtain always the index of the nearest neighbor. But we have noticed that the result is more often the right one with the quantization tree than with the principal axis tree. 
\par For instance, in dimension $2$, on a $5000$ points Gaussian data set, on a million Gaussian query points, we notices: 
\begin{itemize}
\item $56$ percent of false results with PAT.
\item $16$ percent of false results with the quantization tree.
\end{itemize}
\par \noindent Similar results are obtained with other values of the parameters and other data set distributions. This empirical test makes us reasonably optimistic about the performance of a branch and bound tree based on this decomposition. 
\par Still, the cost of the way through the search tree is more expansive with the quantization tree (as described above).
\begin{itemize}
\item For the ``quantization tree'', we have to perform trivial nearest neighbor search to find the right child node. 
\item For ``principal axis tree'', we only compute a projection and perform a binary search. 
\end{itemize}
\par Moreover, it was proved in \cite{Tarpey_PCA_principal_points} that in the case of Gaussian distributions, the affine subspace spanned by stationary quantizers correspond to the first principal components of the considered Gaussian distribution. (This result, extended to the infinite dimensional case in \cite{LuschgyPagesFunctional3} allows to efficiently compute optimal quadratic quantizers of bi-measurable Gaussian processes.) Hence, in this case, this shows that the quantization tree with two branches at each generation is related to the principal axis tree.
\vspace{2mm}
\par \noindent \textbf{First elimination condition} If the center of the Voronoi cell corresponding to the current node is $A$, the first rough method to decide whether a brother node with center $B$ has to be explored or not is compute the distance $d_2$ of the query point $Q$ to the Leibniz halfspace $H(B,A)$. Then the node corresponding to point $B$ is explored if $d_2$ is smaller than the distance to the nearest neighbor found so far, $d_1$. 
We have $d_2 = \frac{AB}{2} - AQ \cos \alpha$ and $QB^2 = QA^2 + AB^2 - 2 AQ AB \cos \alpha$ so that $\Rightarrow \cos \alpha = \frac{QA^2+AB^2-QB^2}{2 AQ AB}$. This yields $d_2 = \frac{QB^2-QA^2}{2 AB}$. Hence, the computation of the distance to the Leibniz halfspace requires one subtractions $QA^2-QB^2$, ($QA^2$ and $QB^2$ can be computed during the search in the quantizer in the parent node), and one multiplication by $\frac{1}{2AB}$. ($\frac{1}{2AB}$ can be computed during the preprocessing.)
\vspace{2mm}
\par \noindent Then, it is clear that the nearest brother node correspond to the second nearest neighbor in the quantizer, and the second nearest to the third nearest neighbor, and so on. Hence, brother nodes have to be explored in the order defined by the distances of its centers the query point. 
\vspace{2mm}
\par
\begin{figure}[!ht]
	\begin{center}
	\psfrag{flatex_d}{\footnotesize{$d$}}
	\psfrag{flatex_d_1}{\footnotesize{$d_1$}}
	\psfrag{flatex_d_2}{\footnotesize{$d_2$}}
	\psfrag{flatex_d_3}{\footnotesize{$d_3$}}
	\includegraphics[width=6cm]{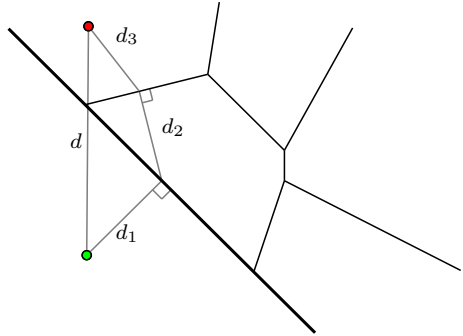}
	\caption{Optimization of the elimination condition for the quantization tree $d^2 \geq d_1^2 +d_2^2+d_3^2$.}
	\label{fig:lowerbound_minimization}
	\end{center}
\end{figure}

 We can also use the same optimization of the lower bound proposed by McNames in \cite{McNamesPAT} and presented in Section \ref{sec:PAT}. Referring to Figure \ref{fig:lowerbound_minimization}, the lower bounds $d_i$ are recursively incremented when exploring brother nodes. 
\par \noindent \textbf{Performance of this first quantization tree algorithm.} This first algorithm has been implemented and its empirical performances has been compared to the two previously exposed PAT and Kd-tree in terms of empirical performances. 
\par Intermediate performances between our implementations of Kd-tree and PAT were obtained in small dimensions. Although, as we will see further in empirical tests, it seems to take better the increase of dimensionality. The preprocessing time, that requires small quantizer computations is also more costly than both PAT and Kd-tree. 
\subsection{Optimizations for the quantization tree}
\par \noindent To reduce the average query time, we are now proposing a new optimization procedure which reduces the number of brother nodes to be checked. 
\vspace{2mm}
\begin{figure}[!ht]
	\begin{center}
	\psfrag{flatex_A}{\small{$A$}}
	\psfrag{flatex_B}{\small{$B$}}
	\psfrag{$-2$}{\small{$-2$}}	
	\psfrag{$-1$}{\small{$-1$}}
	\psfrag{$0$}{\small{$0$}}
	\psfrag{$1$}{\small{$1$}}
	\psfrag{$2$}{\small{$2$}}
	\includegraphics[width=5cm]{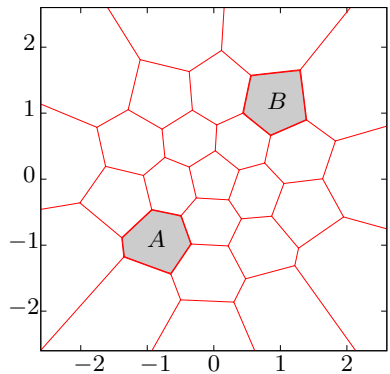}
	\caption{Cell $B$ is ``hidden'' from cell $A$.}
	\label{fig:hidden_cell}
	\end{center}
\end{figure}
\par \noindent Let us consider the Voronoi diagram plotted in Figure \ref{fig:hidden_cell}. In this figure, we obviously know that when the query point is in a cell $A$, its nearest neighbor cannot be in cell $B$, because cell $B$ is ``hidden'' by closer cells. One has to give a precise mathematical sense to ``hidden'' in this sentence. However, in the quantization tree as it has been described, the distance of query point to $H(a,b)$ has to be computed.
\par A first idea is to compute for each $1\leq i \leq n_c$ a list of ``friends'' among brother nodes in which the nearest neighbor can be when $q$ is in cell $i$.
\par This list has to be large enough to ensure that it contains the nearest neighbor but as small as possible in order to reduce the computations of elimination conditions.
\par As concerns the choice of the parameter $n_c$, we have to take in consideration that increasing $n_c$ makes the depth of the tree smaller but also makes the nearest neighbor search slower for each generation of the search tree. 
\vspace{2mm}
\par \noindent \textbf{How can we obtain a friend Voronoi slabs list?} The first observation about obtaining such a friend list is that it is not a simple problem. Indeed, this list is a priori not reduced to slabs whose corresponding Voronoi cells are adjacent in the Voronoi diagram. Moreover, in some cases, the minimal friend list can be quiet large. So is the case for unbounded Voronoi slabs for example. 

\begin{figure}[!ht]
	\begin{center}
	\begin{minipage}[c]{.46\linewidth}
	\psfrag{flatex_A}{$q$}
	\psfrag{flatex_B}{$p$}
	\psfrag{$-2$}{\small{$-2$}}	
	\psfrag{$-1.5$}{\small{$-1.5$}}	
	\psfrag{$-1$}{\small{$-1$}}
	\psfrag{$-0.5$}{\small{$-0.5$}}		
	\psfrag{$0$}{\small{$0$}}
	\psfrag{$0.5$}{\small{$0.5$}}		
	\psfrag{$1$}{\small{$1$}}
	\psfrag{$1.5$}{\small{$1.5$}}	
	\psfrag{$2$}{\small{$2$}}
	\includegraphics[height=65mm]{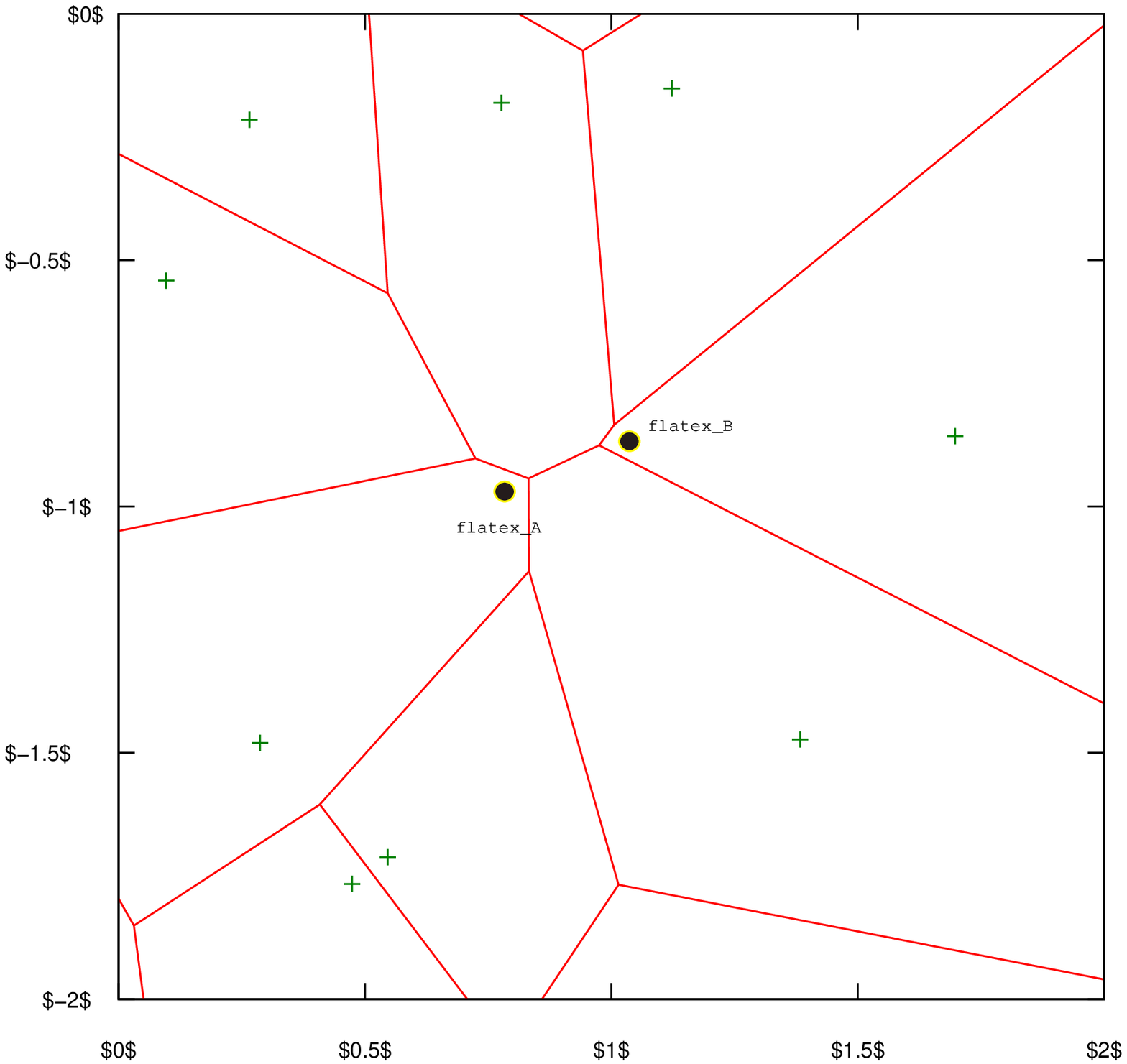}
	\end{minipage}
	\begin{minipage}[c]{.46\linewidth}
	\psfrag{flatex_A}{$q$}
	\psfrag{flatex_B}{$p$}
	\psfrag{$-10$}{\small{$-10$}}
	\psfrag{$-5$}{\small{$-5$}}
	\psfrag{$-4$}{\small{$-4$}}
	\psfrag{$-2$}{\small{$-2$}}	
	\psfrag{$0$}{\small{$0$}}
	\psfrag{$2$}{\small{$2$}}
	\psfrag{$4$}{\small{$4$}}
	\psfrag{$5$}{\small{$5$}}
	\psfrag{$6$}{\small{$6$}}
	\psfrag{$8$}{\small{$8$}}
	\psfrag{$10$}{\small{$10$}}
	\psfrag{$12$}{\small{$12$}}
	\psfrag{$14$}{\small{$14$}}
	\includegraphics[height=65mm]{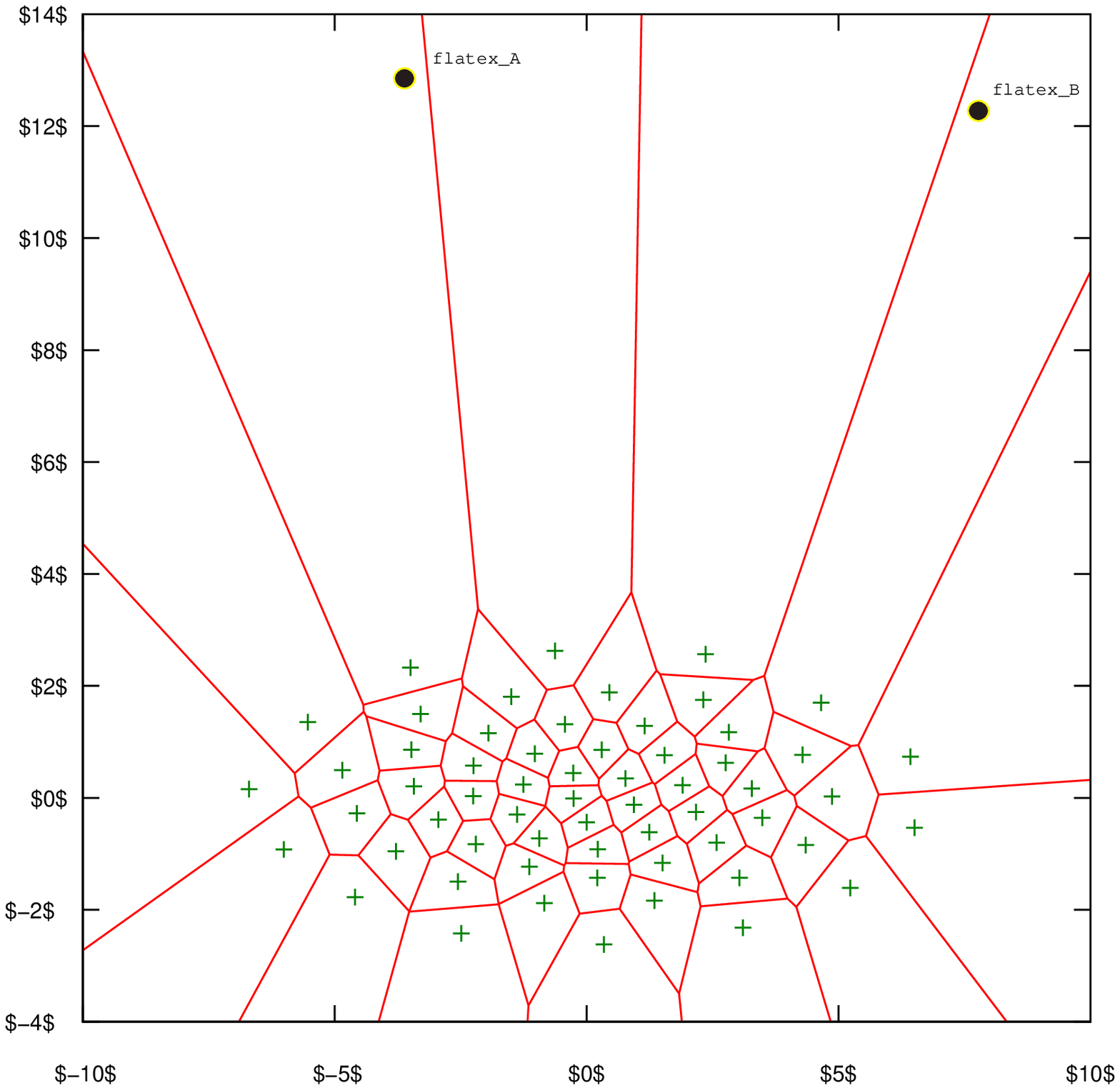}
	\end{minipage}
	\caption{In these cases, the nearest neighbor of the query point $q$ may be $p$ although $p$ is not in an adjacent Voronoi cell.}
	\end{center}
\end{figure}
\section{Some optimizations for the quantization tree algorithm}\label{sec:friend_node_algo}
\par In Section \ref{sec:computational_geometry}, basic definitions about Voronoi diagrams and Delaunay triangulations that are prerequisites to this section ahev been recalled.
\begin{remark}[Voronoi slabs and Voronoi cells]
\par From their respective definitions, one can easily deduce the following properties:
\begin{itemize}
\item Let $S\subset E$ be a finite set of sites, let $C$ be an associated Voronoi partition and consider $s\in S$. Then it is clear that $V(\{s\}) = \ \stackrel{\circ}{\wideparen{\slab_C(s)}}$. 
\item The points of the Voronoi cells $V(R)$ with $R\subset S$ and $\card{R}>1$ belong to the boundaries of Voronoi slabs. 
\item As a consequence, for $s\in S$, as the boundary $V(\{s\})$ is constituted with its faces of lower dimensions, previous remark yields $\overline{V(\{s\})} = \ \overline{\slab(s)}$ and $\delta\slab_S(s) = \partial V_S(\{s\})$.
\end{itemize}
\end{remark}
\par \noindent \textbf{Notations:} In the following of this section, if $S \subset E$ is a finite set of sites in $E$, one will denote $T_S$ the Delaunay triangulation of $S$, $DG_S$ the Delaunay graph of $S$, $V_S$ its Voronoi diagram. For $R\subset S$, $V_S(R)$ will represent the Voronoi cell of $R$ in $S$. If $C_S$ is a Voronoi partition associated to $S$, and $s\in S$, $\slab_S(s)$ will denote the Voronoi slab associated to $S$ is the Voronoi partition $C$. 
\begin{defi}[Leibniz halfspace]
\par For $(a,b) \in E^2$ let us denote $H(a,b) := \Big\{ x \in \R^d | |x-a| \leq |x-b| \Big\}$ the Leibniz halfspace associated to $(a,b)$. 
\end{defi}
\begin{prop}\label{eq:leibniz_inclusion}
\par An obvious property is if $S$ is a finite set of sites of $E$, and $p\in S$, 
$$
V_S(\{p\}) = \bigcap_{s\in S, s\neq p} H(p,s).
$$
\end{prop}
\begin{prop}
If $S$ is a finite set of sites of $E$, and $p\in S$, $V_S(\{p\}) = \bigcap\limits_{\{s,p\}\in DG_S} H(p,s)$.
\end{prop}
\begin{lemm}\label{eq:fundamental_lemma}
\par Let $S\subset E$ be a nonempty finite set of sites in $E$ and $x\in E \backslash S$. Consider $s\in S$, the following assertions are equivalent:
\begin{enumerate}
\item $\{x,s\}\in DG_{S \cup \{x\}}$.
\item $V_S(\{s\}) \cap V_{S\cup \{x\}}(\{x\}) \neq \emptyset$.
\item $V_S(\{s\}) \cap H(x,s) \neq \emptyset$.
\end{enumerate}
\end{lemm}
\par \noindent \textbf{Proof:} \begin{itemize}
\item ($1. \Rightarrow 2.$) Assume that $\{x,s\} \in DG_{S \cup \{x\}}$ then by definition, it is equivalent to $V_{S \cup \{x\}}(\{x,s\}) \neq \emptyset$. 
\par $V_{S \cup \{x\}}(\{x,s\})$ is $(d-1)$-face of $V_{S \cup \{x\}}(x)$. Moreover, by definition of Voronoi cells, $V_{S \cup \{x\}}(\{x,s\}) \subset V_S(\{s\})$, which is open. As a consequence, $\forall y \in V_{S \cup \{x\}}(\{x,s\}), \ \forall \varepsilon >0, B(y,\varepsilon) \cap V_{S \cup \{x\}}(x) \neq \emptyset$. And for small enough $\varepsilon$, $B(y,\varepsilon) \subset V_S(\{s\})$. We can conclude that $V_S(\{s\}) \cap V_{S\cup \{x\}}(\{x\}) \neq \emptyset$.
\item ($2. \Rightarrow 3.$) is obvious owing to Proposition \ref{eq:leibniz_inclusion}.
\item ($3. \Rightarrow 1.$) If $y \in V_S(\{s\}) \cap H(x,s)$, let us show that $V_{S \cup \{x\}}(\{x,s\}) \neq \emptyset$.
\par Consider the segment $[s,y]$. By convexity, $[s,y] \subset V_S(\{s\})$. Thus every point of $[s,y]$ is closer to $s$ than to any other point of $S$. On the other hand, it can either be closer to $s$ than to $x$, or closer to $x$ than to $s$ or at the same distance. 
\par We now define the applications $f : [0,1] \to [s,y] \subset E$ by $f(\lambda) = \lambda s + (1-\lambda)y$ and $\Delta : E \to \R$ by $\Delta(p) = d(p,x) - d(p,s)$.

\par $\Delta \circ f$ is a continuous function with $\Delta \circ f(0)>0$, $\Delta \circ f(1)<0$. The intermediate value theorem shows that there exists $\lambda^*$ such that $\Delta \circ f(\lambda^*)=0$ and thus $f(\lambda^*) \in V_{S \cup \{x\}}(\{x,s\})$. \myqed
\end{itemize}
\begin{figure}[!ht]
	\begin{center}
	\psfrag{$x$}{$x$}
	\psfrag{$s$}{$s$}
	\psfrag{$V_S(s)$}{$V_S(\{s\})$}
	\psfrag{$H(x,s)$}{$H(x,s)$}
	\psfrag{$inter$}{$H(x,s) \cap V_S(\{s\})$}
	\psfrag{$vxs$}{$V_{S \cup \{x\}}(\{x,s\})$}
	\psfrag{$-1$}{\small{$-1$}}
	\psfrag{$-0.5$}{\small{$-0.5$}}		
	\psfrag{$0$}{\small{$0$}}
	\psfrag{$0.5$}{\small{$0.5$}}		
	\psfrag{$1$}{\small{$1$}}
	\psfrag{$1.5$}{\small{$1.5$}}	
	\psfrag{$2$}{\small{$2$}}
	\includegraphics[width=7cm]{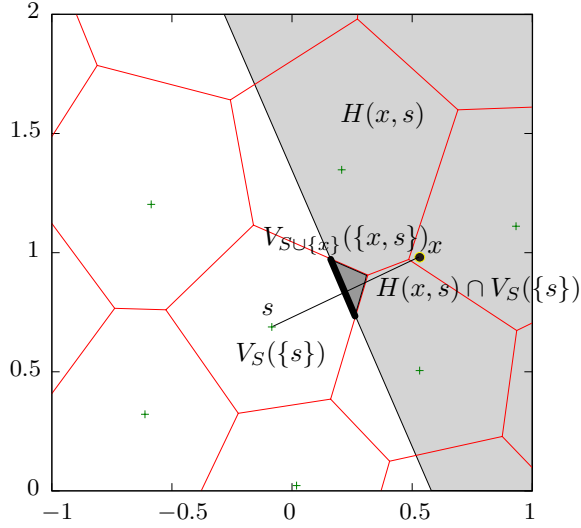}
	\caption{If the query point $q$ lies on the dark grey region $H(x,s) \cap V_S(\{s\})$ its nearest neighbor may be $x$.}
	\end{center}
\end{figure}
\par The first modification made in the quantization tree algorithm is to assume that the points of the quantizer at each generation are points of the underlying codebook $\Gamma$. (In order to fulfill this requirement, we project an optimal quantizer onto the codebook.)
\begin{coro}
\par Let $\Gamma= \{\Gamma_1,\cdots,\Gamma_n\}$ be a codebook of $E$. $S = \{s_1,\cdots,s_p\} \subsetneq \Gamma$ be subset of $\Gamma$. Let $\Proj_\Gamma$ be a nearest neighbor projection on $\Gamma$. $\Gamma$ is being partitioned into $p$ subsets $\Gamma^1,\cdots,\Gamma^p$ with $\Gamma_i= \Gamma \cap \slab_S(s_i)$, by their nearest neighbor projection on $S$.
\par \noindent Consider $q \in E$. If $q\in \slab_S(s)$ and $t=\Proj_\Gamma(s)$ then $\{t,s\} \in DG_{S \cup \{t\}}$.
\end{coro}
\par \noindent \textbf{Proof:} This is a straightforward consequence of the previous lemma. \myqed
\vspace{2mm}
\par \noindent \textbf{Notation:} Let $S$ be a set of sites in $E$. For a point $t$ in $E$, we denote $PI_S(t) = \Big\{ s\in S, \{s,t\} \in DG_{S \cup \{t\}} \Big\}$. The notation $PI$ stands for ``Pseudo-Insertion''. 
\par From an algorithmic viewpoint, the Delaunay graph of $S$ being computed, $PI_S(t)$ stands for the sets of points in $S$, that are connected to $t$ when updating the Delaunay graph to take account of this new point. 
\par Implementing a procedure that computes $PI_S(t)$ is very similar to the insertion procedure of point $t$ in $T_S$.
\par \noindent \textbf{First friend node algorithm:} This leads to a first method to compute a friend list:
\begin{breakbox}
\par \noindent For every point $p$ of the underlying codebook,
\begin{itemize}
\item Compute $s= \Proj_S(p)$ and $PI_S(p)$. 
\item Then for every point $s'\in PI_S(p)$, insert $s$ in the set of friends of node $s'$.
\end{itemize}
\end{breakbox}
\par This method gives a first algorithm to compute friend list. Still, when the data set is large, it is very expensive because one has to deal with all the points of the data set. 
\par In fact it is possible to compute an acceptable friend list thanks to the same result \ref{eq:fundamental_lemma} without using the points of the underlying data set. 
\vspace{2mm}
\par \noindent \textbf{Fast friend node algorithm:} In this section, another method to compute friend node lists is devised which does not need to deal with the complete underlying data set but only the underlying codebook.
\par When keeping the same notations, the principle of the method is to compute for every $\slab_S(s)$, $s\in S$ ,of the Voronoi partition $C_S$, the set $UPI_S(s) := \!\!\!\!\!\! \bigcup\limits_{p \in \slab_S(s)} \!\!\!\!\!\! PI_S(p)$. It is the union of all the pseudo-insertions of points of $\slab_S(s)$. If one is able to compute this set, the resulting friend nodes algorithm simply writes:
\vspace{2mm}
\begin{breakbox}
For every point $s \in S$,
\begin{itemize}
\item Compute $UPI_S(s)$. 
\item Then for every point $s'\in UPI_S(s)$, insert $s$ in the set of friends of node $s'$.
\end{itemize}
\end{breakbox}
\par \noindent The question is: how can we compute $UPI_S(s)$?
\begin{lemm}
\par With the same notations, one has $UPI_S(s) = \!\!\!\!\!\! \bigcup\limits_{p \in \delta\slab_S(s)} \!\!\!\!\!\! PI_S(p)$. In other words, we have to check points of the boundary $\delta\slab_S(s)$ of $\slab_S(s)$. 
\end{lemm}
\begin{remark}
\par Let us recall that, thanks to Proposition \ref{thm:slab_and_cells}, $(\delta\slab_S(s) = \partial V_S(\{s\})$.
\end{remark}
\par \noindent \textbf{Proof:} Consider $x \in \slab_S(s)$ such as $s' \in PI_S(x)$. Let us define $x^*$, such that $\{ x^* \} = [x,s'] \cap \partial V_S(s)$.
\begin{itemize} 
\item One has $H(x^*,s') \supset H(x,s')$. So $V_S(\{s'\}) \cap H(x^*,s') \supset V_S(\{s'\}) \cap H(x,s')$, hence $V_S(\{s'\}) \cap H(x,s') \neq \emptyset \Rightarrow V_S(\{s'\}) \cap H(x^*,s') \neq \emptyset$ that is equivalent to $s' \in PI(x^*)$ thanks to the Lemma \ref{eq:fundamental_lemma}.
\item Finally, $\forall x\in \slab_S(s), \forall s'\in PI_S(x), \exists x^* \in \delta\slab_S(s) \textrm{ such that } s'\in PI_S(x^*)$. \myqed
\end{itemize}
\begin{remark}
\par As there are not a finite number of sites on the boundaries, this does not give an effective method for computing $UPI_S(s)$ yet. 
\end{remark}
\par \noindent As seen in Section \ref{sec:voronoi_delaunay_algo}, computing the set $PI_S(x)$ corresponds almost to the same algorithm as the insertion procedure in an incremental triangulation algorithm, that is:
\begin{itemize}
\item Localization of $x$ in the triangulation,
\item Computation of the set $ICL(x)$,
\item $UI_S(x)$ is the set of points that belong to a cell of $ICL(x)$ plus, if $x$ is outside the convex hull of $S$, the points of the external faces of $T_S$ that are visible from $x$.
\end{itemize}
\begin{lemm}
\par Let $S$ be a non empty finite set of sites in $E$. We consider the circumsphere $C$ of Delaunay $d$-cell of the Delaunay triangulation $T_S$. We denote $c$ its center and $r$ its radius. Let $s$ be a site of $S$. 
\par \noindent If $V_S(\{s\}) \subset C \neq \emptyset$ then $c+\frac{r}{|s-c|}(s-c) \in V_S(\{s\})$.
\end{lemm}
\par \noindent The proof is straightforward. This leads to an algorithm to compute sets $(UPI_S(s))_{s \in S}$. 
\begin{breakbox}
	\begin{itemize}
	\item For every Delaunay $d$-cell $D$ of $T_S$ 
		\begin{itemize}
		\item Compute the center $c$ and radius $r$ of its circumsphere.
		\item For every site $s \in S$ that is not in $D$, compute $p:=c+r\frac{s-c}{|s-c|} \in V_S(\{s\})$, and check if the site $s$ is the nearest neighbor of $p$ in $S$. If so is the case, then the points of the Delaunay $d$-cells $D$ belong to $UPI_S(s)$.
		\end{itemize}
	\item Then deal with unbounded Voronoi cells:
		\begin{itemize}
		\item For every external face $F$ of the Delaunay triangulation, compute a normal vector $u_F$ directed toward the exterior of the convex hull of $S$.
		\item For two distinct external faces $F_1$ and $F_2$ of the Delaunay triangulation, if $\langle u_{F_1},u_{F_2} \rangle >0$ then for every $(s_1,s_2) \in F_1\times F_2$, $s_1 \in UPI_S(s_2)$ and $s_2 \in UPI_S(s_1)$.
		\end{itemize}
	\end{itemize}
\end{breakbox}
\vspace{2mm}
\par
\begin{figure}[!ht]
	\begin{center}
	\begin{minipage}[c]{.46\linewidth}
	\psfrag{-4}{$-4$}
	\psfrag{-3}{$-3$}
	\psfrag{-2}{$-2$}
	\psfrag{-1}{$-1$}
	\psfrag{0}{$0$}
	\psfrag{1}{$1$}
	\psfrag{2}{$2$}
	\psfrag{3}{$3$}
	\psfrag{4}{$4$}
	\includegraphics[height=65mm]{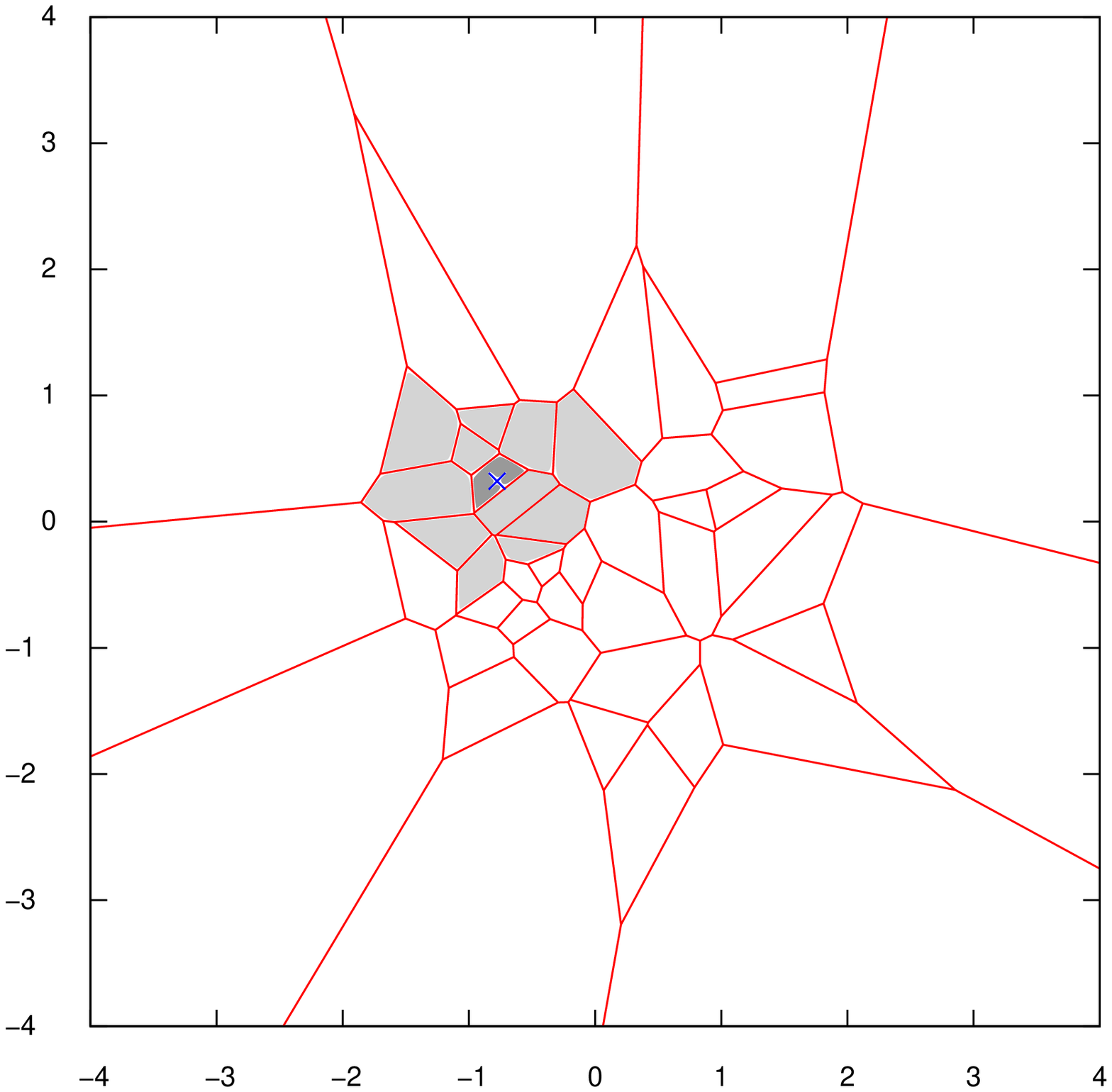}
	\end{minipage}
	\begin{minipage}[c]{.46\linewidth}
	\psfrag{-4}{$-4$}
	\psfrag{-3}{$-3$}
	\psfrag{-2}{$-2$}
	\psfrag{-1}{$-1$}
	\psfrag{0}{$0$}
	\psfrag{1}{$1$}
	\psfrag{2}{$2$}
	\psfrag{3}{$3$}
	\psfrag{4}{$4$}
	\includegraphics[height=65mm]{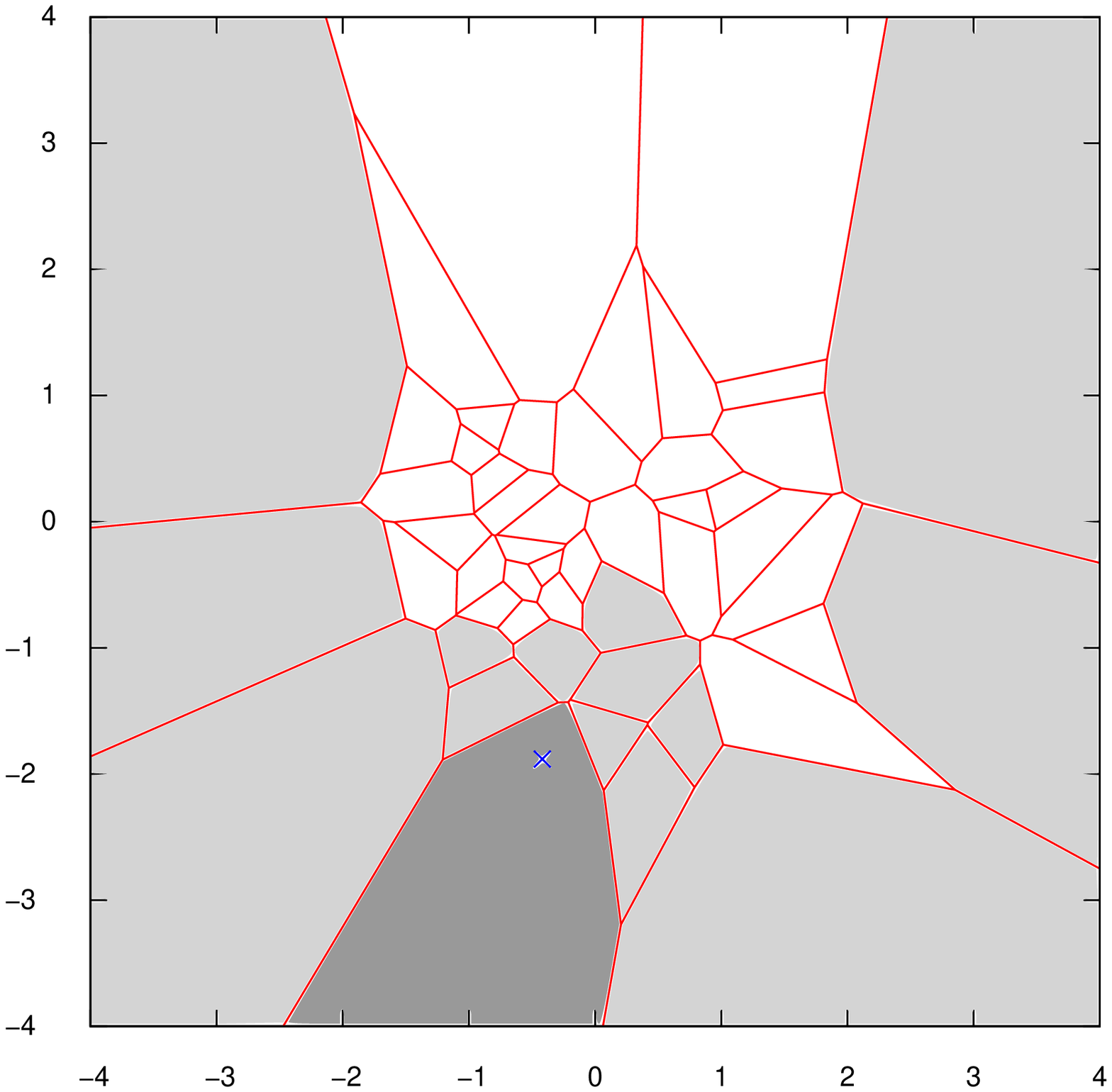}
	\end{minipage}
	\caption{Examples of friend Voronoi cells in a two-dimensional Voronoi diagram in the case of a bounded Voronoi cell (left) and in the unbounded case (right). In both case, the dark gray region is the considered Voronoi cell and the light gray regions are the friend Voronoi cells.}
	\label{fig:friend_examples}
	\end{center}
\end{figure}

 \noindent In Figure \ref{fig:friend_examples}, we present some friend Voronoi lists in the $2$-dimensional case. 
\section{Test with real data sets}
\par To perform the following tests, the quantization tree algorithm and the friend-node optimization have been implemented in C++. Because of the additional feature related to computational geometry that we needed, as the pseudo-insertion procedure, we had to implement a Delaunay triangulation. All the figures presented in this article were generated with this implementation of the Voronoi diagram with which we performed the following tests. 
\subsection{Tests on Gaussian and uniform data sets}
\par In Tables \ref{tab:test_on_gaussian_distribution1}, \ref{tab:test_on_gaussian_distribution2} and \ref{tab:test_on_unfirom_distribution}, we report the execution time for $10$ millions nearest neighbor queries on data-sets of size $5000$ generated with independent Gaussian pseudo random variables and with a uniform distribution on the hypercube. The best overall performances were obtained with $n_c = 35$ children by node for the quantization tree. The tests were performed with an Intel Pentium Dual CPU at $2$GHz. We noticed that in dimension $d=2$ and $d=3$, we had intermediate performances between the ``principal axis tree'' and the Kd-tree algorithms. In dimension $4$, the performance of the ``principal axis tree'' and the ``quantization tree'' are close one to each other. Finally, it seems that the quantization tree has a better behaviour in dimensions greater than $5$ where it significantly outperforms the two other implemented methods.

\begin{figure}[!ht]
\begin{center}
{
\begin{tabular}{|c|c|c|c|c|c|c|c|}
\hline
					& $d = 2$	& $d = 3$	& $d = 4$	& $d = 5$	& $d = 6$	& $d = 7$	& $d=8$\\
\hline \hline
Quantization tree	& $1.76$s	& $2.75$s	& $5.35$s	& $8.93$s	& $15.99$s	& $28.06$s	& $52.31$s\\
\hline
Principal axis tree	& $1.21$s	& $1.86$s	& $4.49$s	& $10.87$s	& $20.14$s	& $41.56$s	& $82.30$s\\
\hline
Kd-tree				& $1.88$s	& $3.71$s	& $8.54$s	& $17.13$s	& $31.06$s	& $60.67$s	& $118.93$s\\
\hline
\end{tabular}
\caption{Execution time of $10$ millions random queries on a data set of $5000$ points, generated with a Gaussian pseudo random generator.}
\label{tab:test_on_gaussian_distribution1}
}
\end{center}
\end{figure}

\begin{figure}[!ht]
\begin{center}
{
\begin{tabular}{|c|c|c|c|c|c|c|c|}
\hline
					& $d = 2$	& $d = 3$	& $d = 4$	& $d = 5$	& $d = 6$	& $d = 7$	& $d=8$\\
\hline \hline
Quantization tree	& $2.59$s	& $3.87$s	& $6.46$s	& $11.90$s	& $27.54$s	& $45.78$s	& $84.63$s\\
\hline
Principal axis tree	& $1.33$s	& $2.44$s	& $4.94$s	& $12.78$s	& $41.02$s	& $62.33$s	& $119.88$s\\
\hline
Kd-tree				& $2.82$s	& $5.20$s	& $11.32$s	& $24.20$s	& $47.51$s	& $87.61$s	& $164.52$s\\
\hline
\end{tabular}
\caption{Execution time of $10$ millions random queries on a data set of $10000$ points, generated with a Gaussian pseudo random generator.}
\label{tab:test_on_gaussian_distribution2}
}
\end{center}
\end{figure}

\begin{figure}[!ht]
\begin{center}
{
\begin{tabular}{|c|c|c|c|c|c|c|c|}
\hline
					& $d = 2$	& $d = 3$	& $d = 4$	& $d = 5$	& $d = 6$	& $d = 7$	& $d=8$\\
\hline \hline
Quantization tree	& $1.62$s	& $2.30$s	& $3.75$s	& $6.47$s	& $10.33$s	& $15.91$s	& $32.62$s\\
\hline
Principal axis tree	& $0.74$s	& $1.52$s	& $2.81$s	& $6.71$s	& $16.53$s	& $28.03$s	& $47.53$s\\
\hline
Kd-tree				& $1.54$s	& $2.82$s	& $5.46$s	& $10.64$s	& $18.50$s	& $31.60$s	& $55.71$s\\
\hline
\end{tabular}
\caption{Execution time of $10$ millions random queries on a data set of $5000$ points, generated with a uniform pseudo random generator.}
\label{tab:test_on_unfirom_distribution}
}
\end{center}
\end{figure}

\begin{remark}[Computational cost or the preprocessing for the friend cell algorithm]
\par An important fact that we have experienced is that, in higher dimensions, the friend cells list becomes bigger and there is no more competitive advantage in using it in dimension higher than $7$ (when having less than $30$ branches per generation in the quantization tree). Moreover, as it requires to compute Delaunay triangulations during the preprocessing, whose complexity exponentially increases with the dimension, the computational cost of the friend cell preprocessing makes it useless in higher dimensions. 
\end{remark}

\vspace{5mm}
\par \noindent The author is very grateful to Gilles Pagès (LPMA University Paris VI) for his helpful remarks and comments, and to Johan Mabille (Natixis) for his advices concerning the practical implementation. 
\bibliography{biblio}

\end{document}